\documentclass{article}

\usepackage{PRIMEarxiv}
\usepackage{multirow}
\usepackage[utf8]{inputenc} 
\usepackage[T1]{fontenc}    
\usepackage{hyperref}       
\usepackage{url}            
\usepackage{booktabs}       
\usepackage{amsfonts}       
\usepackage{nicefrac}       
\usepackage{microtype}      
\usepackage{lipsum}
\usepackage{fancyhdr}       
\usepackage{graphicx}       
\usepackage{subcaption}
\graphicspath{{media/}}     

\title{AutoPET III Challenge: Tumor Lesion Segmentation using ResEnc-Model Ensemble
}

\author{
  Tanya Chutani, Saikiran Bonthu, Pranab Samanta, Nitin Singhal \\
  Airamatrix Pvt Ltd \\
  Mumbai 400604 \\
  India. \\
  \texttt {solutions@airamatrix.com} \\
}

\begin{document}
\maketitle

\begin{abstract}
Positron Emission Tomography (PET) /Computed Tomography (CT) is crucial for diagnosing, managing, and planning treatment for various cancers. Developing reliable deep learning models for the segmentation of tumor lesions in PET/CT scans in a multitracer multicenter environment, is a critical area of research. Different tracers, such as Fluorodeoxyglucose (FDG) and Prostate-Specific Membrane Antigen (PSMA), have distinct physiological uptake patterns and data from different centers often vary in terms of acquisition protocols, scanner types, and patient populations. Because of this variability, it becomes more difficult to design reliable segmentation algorithms and generalization techniques due to variations in image quality and lesion detectability. To address this challenge, We trained a 3D Residual encoder U-Net within the nnU-Net framework, aiming to generalize the performance of automatic lesion segmentation of whole body PET/CT scans, across different tracers and clinical sites. Further, We explored several preprocessing techniques and ultimately settled on using the ‘TotalSegmentator’ to crop our training data. Additionally, we applied resampling during this process. During inference, we leveraged test-time augmentations and other post-processing techniques to enhance tumor lesion segmentation. Our team currently hold the top position in the AutoPET III challenge and outperformed the challenge baseline model in the preliminary test set with Dice score of 0.9627. Github link: \url {https://github.com/tanya-chutani-aira/autopetiii}

\end{abstract}
\section{Introduction}
Over the past decades, PET/CT \cite{autopet,petct} has become a crucial tool in oncological diagnostics, management, and treatment planning. In clinical practice, medical experts usually depend on qualitative analysis of PET/CT images, although quantitative analysis could provide more precise and individualized tumor characterization and therapeutic decisions. A significant barrier to clinical adoption is lesion segmentation, a necessary step for quantitative image analysis. When done manually, it is tedious, time-consuming, and costly. Machine learning offers the potential for rapid and fully automated quantitative analysis of PET/CT images. 

\section{Methodology}

\begin{figure}
\centering
  \includegraphics[width=1.0\linewidth]{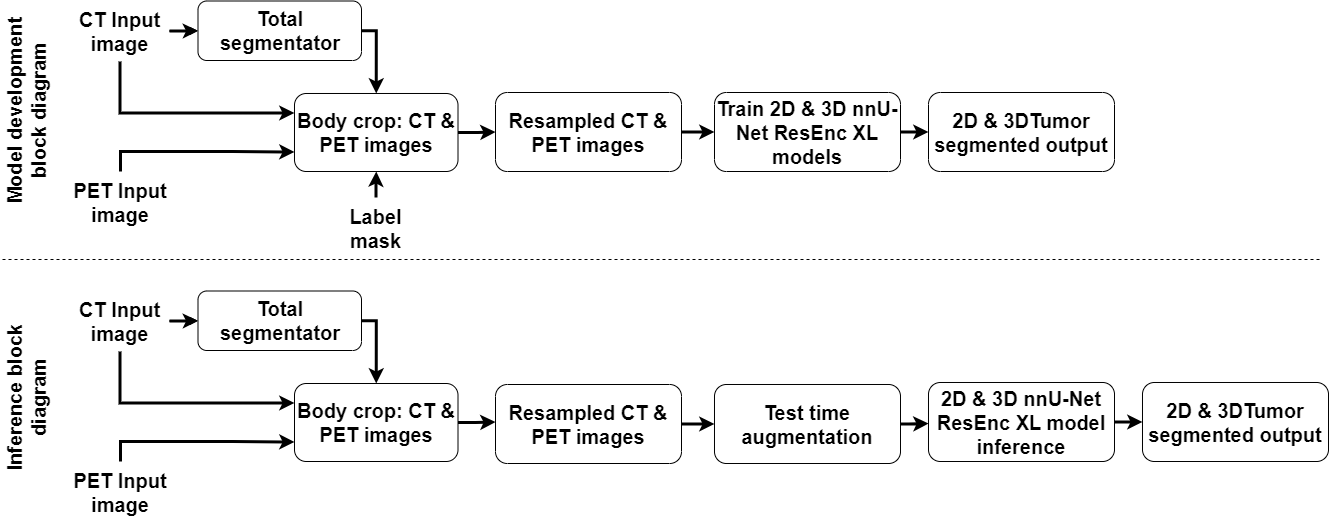}
    \caption{Block diagram for model development and model inference}
    \label{fig:architecture}
\end{figure}

The block diagram in Fig. \ref{fig:architecture} illustrates the training and inference pipeline utilized for the AutoPET III challenge.

\subsection{Development dataset}
The AutoPET III challenge dataset comprises 1,611 images in the training set. Specifically, it includes 1,014 FDG \cite{fdgpsma} images from 900 patients and 597 PSMA images \cite{psma-differentballgame} from 378 patients. The validation set is derived from 5 images (3 FDG images and 2 PSMA images). The final evaluation will use a hidden test set of 200 images, with half of the test data matching the sources and distributions of the training data, while the other half will be cross-sourced from the opposite center, ensuring robust model assessment.
The FDG PET/CT training and testing images from UKT hospital were annotated by a radiologist with a decade of experience in hybrid imaging. The PSMA PET/CT datasets from both centers (UKT and LMU hospital) underwent initial annotation by a single reader and subsequent review by a radiologist with five years of experience in hybrid imaging.

\subsection{Data pre-processing}
Training a model for 3D medical images, such as PET/CT scans \cite{petct}, is challenging due to the substantial memory consumption required during both training and inference. Additionally, not all parts of the image contain relevant structures (e.g., tumors), and the image boundaries often have noise or artifacts. 

Data pre-processing steps include TotalSegmentator \cite{totalseg}, image resampling, data augmentation, and data normalization. 

\subsubsection{TotalSegmentator:}
To localize the tissue and mitigate the boundary noise, we used TotalSegmentator \cite{totalseg} to crop the 3D PET/CT images. Cropping reduces the image size while retaining relevant information, leading to more memory-efficient training. TotalSegmentator intelligently crops to preserve relevant structures (e.g., tumors) while minimizing unnecessary background and noisy regions, ensuring cleaner input data for the model. This approach ensures uniform input sizes across all samples, making it easier to batch and train models efficiently. 

The trade-off is that aggressive cropping can sometimes sacrifice spatial context to save memory. To balance this, we used overlapping crops during training, ensuring that neighboring context is still considered even if individual crops lose some spatial information. Additionally, we applied padding to the cropped regions, allowing the model to learn from nearby context. To make the generalized model we applied data augmentation module. Data augmentation module is followed by cropped data.

In TotalSegmentator, we utilized a 3D lower nnU-Net to accurately segment the body from the PET/CT image.

\subsubsection{Data augmentation:}
Data augmentation plays a crucial role in enhancing model robustness and generalization. We employ basic transformations such as random flips, random rotations, elastic deformations, intensity scaling which are applied in standard nnU-Netv2 \cite{nnunetv2}.

\subsubsection{Data normalization:}
We applied data normalization to the cropped images, including resampling and intensity normalization, using standard nnUNetv2 plans defined for CT and PET. Resampling ensures consistent spatial resolution across all images, while intensity normalization brings the image values to a common scale, reducing variability caused by different imaging protocols. 

\subsection{Model architecture}
We employed nnU-Net ResEnc XL from nnUNetv2 pipeline as the model architecture for training \cite{nnunetv2,ResencNet}. The model is modified version of the ResNet family network. It utilizes residual block similar to the ResNet [Ref.]. nnU-Net ResEnc XL represents a significant advancement in the field of medical imaging, leveraging deep learning to enhance the accuracy and efficiency of tumor lesion segmentation in 3D PET/CT scans. 

\subsection{Training methodology}
During training, we adopted a robust approach by dividing the data into five folds. For each fold, 80 percentage of the data served as the training set, while the remaining 20 percentage acted as the validation set. This strategy allowed our models to capture variations effectively and ensured their robustness when faced with unseen test data. We explored both 2D and 3D versions of the nnU-Net ResEnc XL architecture and employed a patch size of [224, 192, 224] in the Patch-Based Training of nnU-Netv2 pipeline.

\section{Results and discussion}
\subsection{Experimental setup}
For this experimentation, we have used nnU-Netv2 framework with nnU-Net ResEnc XL architecture. The framework is developed in pytorch and we have used 48GB GPU to develop the 2D and 3D models for each fold.      
\subsection{nnU-Net ResEnc XL results and discussion}
In nnU-Net, we have used 2D and 3D nnU-Net ResEnc XL model in full resolution. We assessed the performance of 2D and 3D segmentation models using a 5-fold cross-validation approach with the Dice metric. The results, presented in Table \ref{tab:my_label}, indicate that 3D models outperformed 2D models in tumor lesion segmentation in PET/CT images when trained on a cropped dataset generated using TotalSegmentator.
 
\begin{table}[h]
    \centering
    \caption{Dice Score in 5-fold cross validation set}
    \begin{tabular}{|c||ccccc|}
         \hline
         Model & Fold-0 & Fold-1 & Fold-2 & Fold-3 & Fold-4\\
         \hline
         \cline{2-6}
          ResEnc XL-2D & 0.5965 & 0.6014 & 0.5706 & 0.6207 & 0.5647 \\
         \hline
         \cline{2-6}
         ResEnc XL-3D & 0.6856 & 0.6450 & 0.6244 & 0.6665 & 0.6171 \\
         \hline
    \end{tabular}
    \label{tab:my_label}
\end{table}

\begin{figure}
     \begin{subfigure}[b]{0.5\textwidth}
         \centering
         \includegraphics[width=\linewidth]{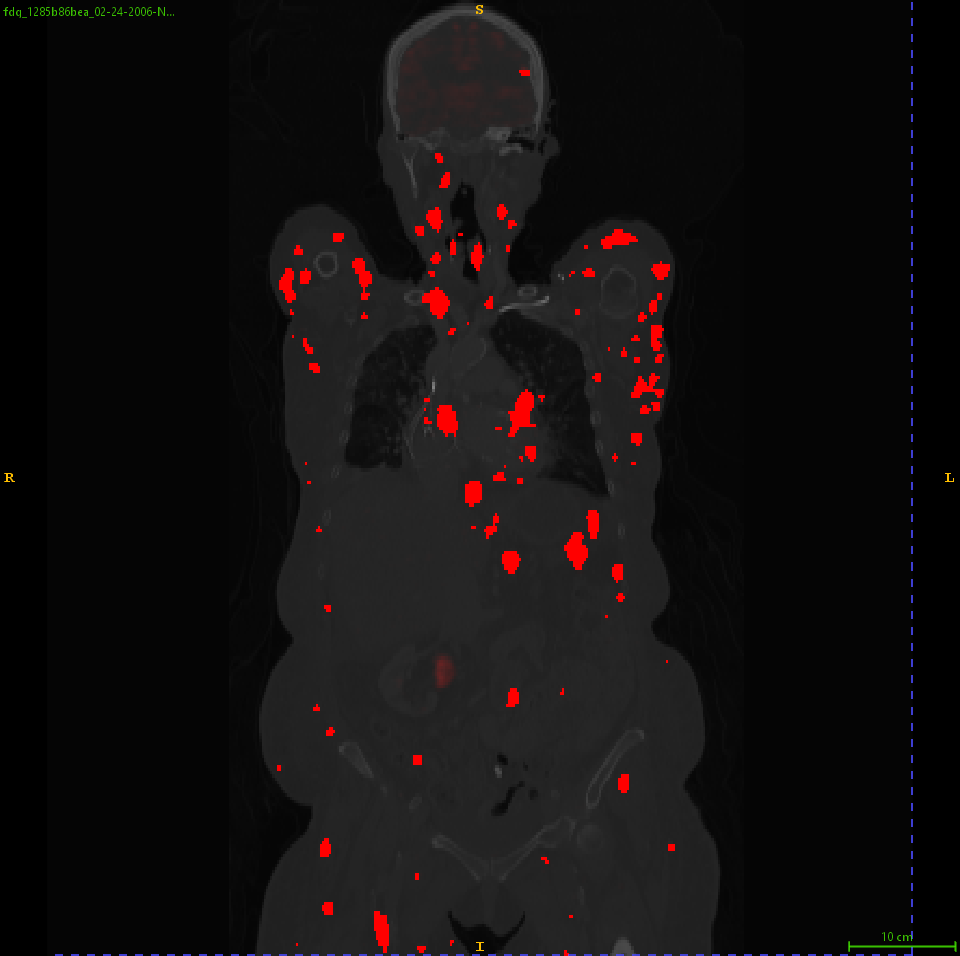}
         \caption{Whole body FDG image with lesion ground truth}
         \label{fig: FDG_gt}
     \end{subfigure}
     \hfill
     \begin{subfigure}[b]{0.5\textwidth}
         \centering
         \includegraphics[width=\linewidth]{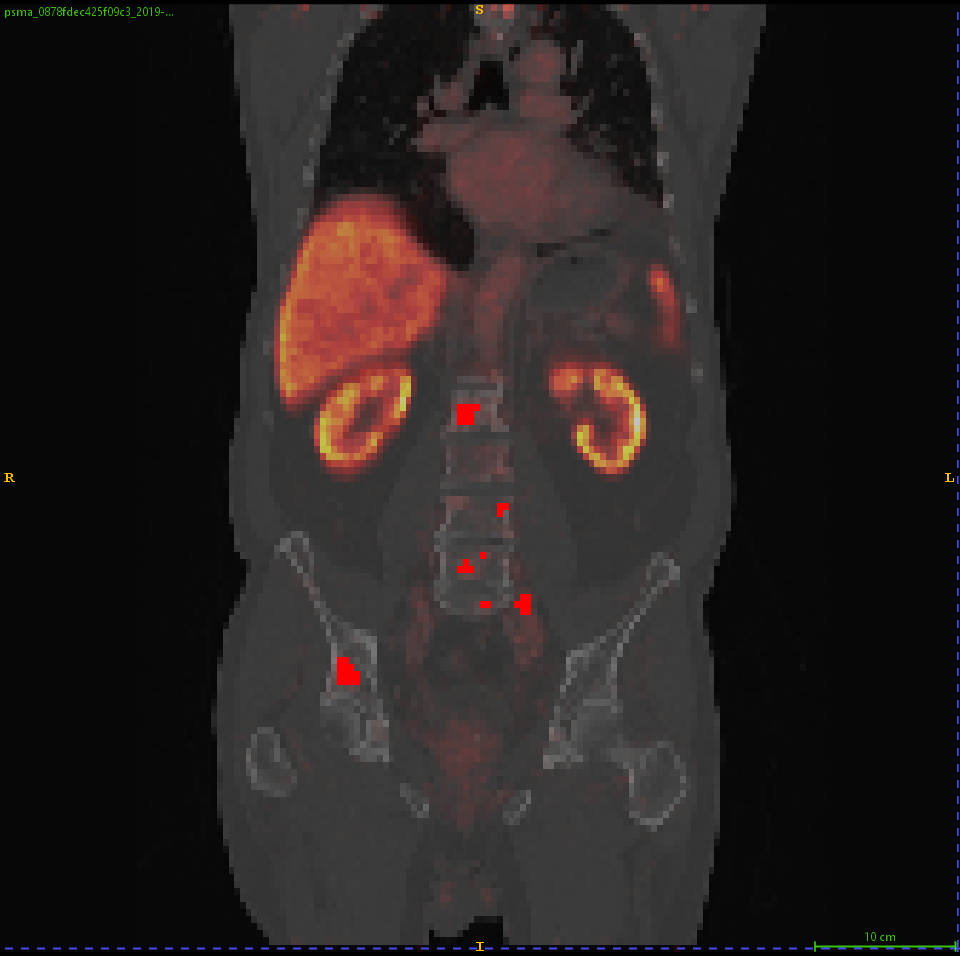}
         \caption{Whole body PSMA image with lesion ground truth}
         \label{fig: PSMA_gt}
     \end{subfigure}
     \hfill
     \hfill
     \begin{subfigure}[b]{0.5\textwidth}
         \centering
         \includegraphics[width=\linewidth]{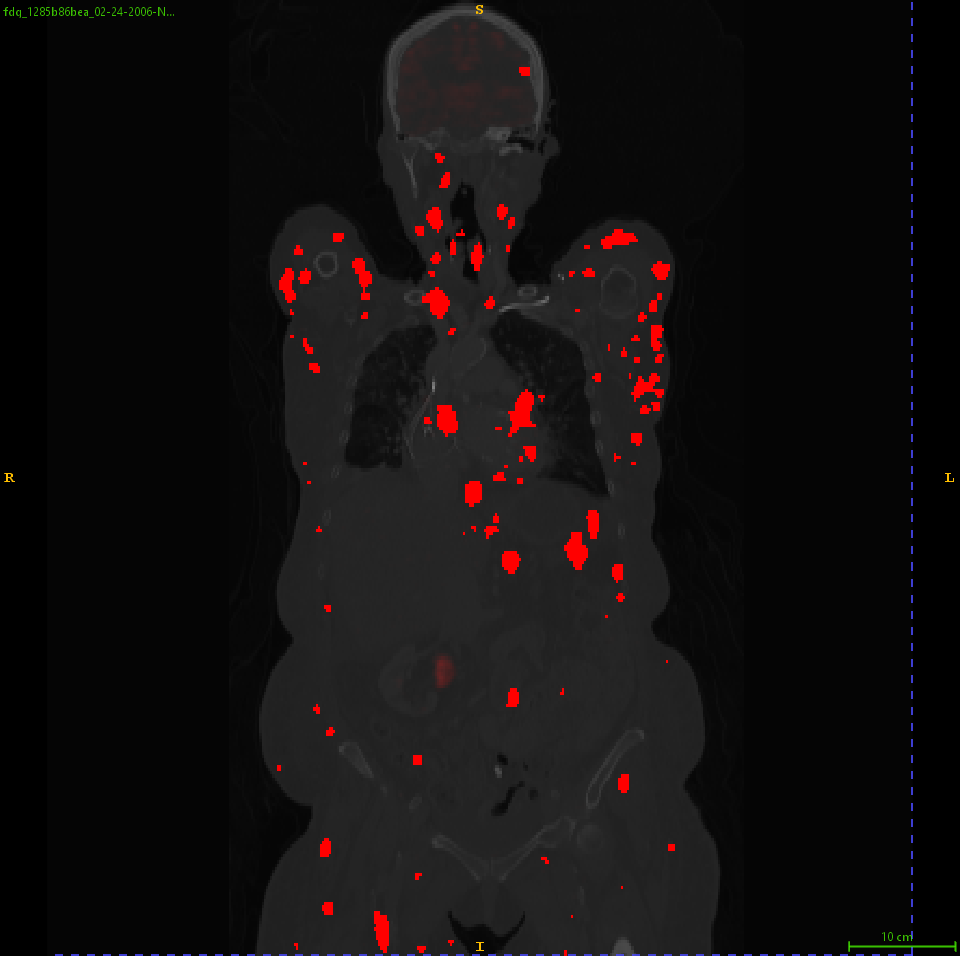}
         \caption{Whole body lesion segmentation results of FDG image}
         \label{fig: FDG_preds}
     \end{subfigure}
     \begin{subfigure}[b]{0.5\textwidth}
         \centering
         \includegraphics[width=\linewidth]{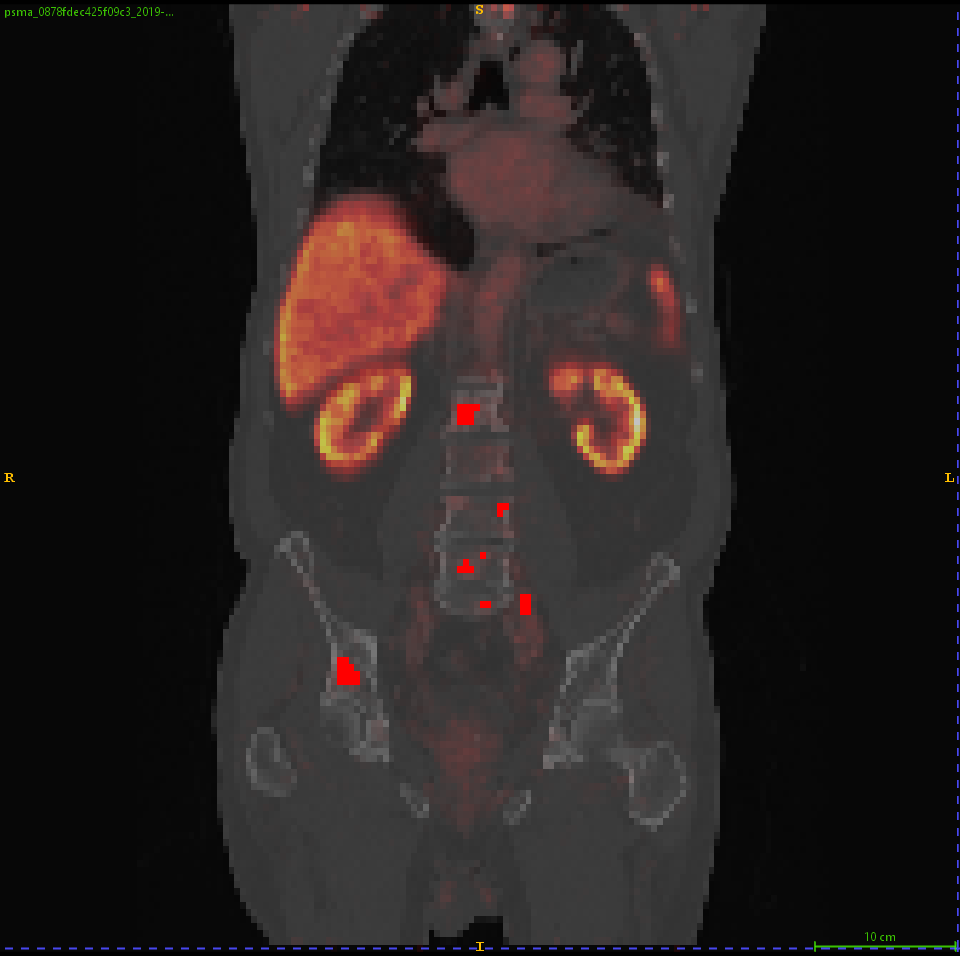}
         \caption{Whole body lesion segmentation results of PSMA image}
         \label{fig: PSMA_preds}
     \end{subfigure}
        \caption{Qualitative results on FDG and PSMA images}
        \label{fig:FDG_PSMA}
\end{figure}

\subsection{Model ensemble results and discussion}
For the AutoPET III challenge, we employed an ensemble of 5-fold models, and the STAPLE algorithm \cite{Staple} to combine the results from the five models. This ensemble approach facilitated the improved generalization, robustness and performance needed for tumor lesion segmentation in a multicenter, multi-tracer environment. 

Integrating models using 5-fold cross-validation presents significant challenges, particularly when faced with time constraints during inference. The computational demands are high, contributing to overall complexity. To address these issues, we implemented specific strategies. First, we optimized CPU usage by preprocessing input images only once. Additionally, we parallelized inference across five different models using a multithreading approach. Furthermore, dynamic test-time augmentations (TTA) were employed during inference to enhance robustness and accuracy in tumor lesion segmentation \cite{TTAaug}. For larger images, we judiciously limited TTA to minimize augmentation time. Our preliminary results on the autoPET III challenge test set validate the effectiveness of this training strategy. 

Upon comparing our predicted segmentations with ground truth annotations, our model exhibits remarkable accuracy, particularly in tumor lesion detection and boundary delineation. Figure \ref{fig:FDG_PSMA} showcases the ground truth of FDG and PSMA, followed by the segmentation results for tumor lesions in whole-body PET/CT images.

\subsection{Model Submissions}
Initially, we submitted three versions of models for the preliminary test set:
\begin{itemize}
    \item A 2D single-fold model achieved a Dice score of 0.9627.
    \item A 2D 5-fold ensemble of models achieved a Dice score of 0.9602.
    \item A 3D 5-fold ensemble of models achieved a Dice score of 0.7530.
\end{itemize}

Moving on to the final test set, we submitted two additional versions of models:
\begin{itemize}
    \item A 2D 3-fold ensemble of models, resulting in a Dice score of 0.84 on the preliminary test set.
    \item A 3D 4-fold ensemble of models, resulting in a Dice score of 0.74 on the preliminary test
\end{itemize}

\section{Conclusions}
To address the challenge of tumor lesion segmentation of PET/CT images from multiple tracers and centers, we proposed a model training approach using a cropped dataset with TotalSegmentator, followed by standard pre-processing steps. However, during inference TotalSegmentator with dynamic test time augmentation helps to improve the overall performance across the 5-fold using 3D models that reflected in the validation leader board as well in AutoPET III challenge. Moreover, 3D model ensemble with STAPLE algorithm outperformed with respect to the individual model performance.

\section{Acknowledgement}
We extend our gratitude to the organizers of the AutoPET III challenge for generously providing the data, enabling us to conduct fascinating research on tumor lesion segmentation in whole-body PET/CT images.

%
%
%
%



\end{document}